\documentclass[showpacs,preprintnumbers,amsmath,amssymb,floatfix]{revtex4-1}

\usepackage{graphicx}
\usepackage{dcolumn}
\usepackage{bm}
\usepackage{amssymb}
\usepackage{epsfig}
\usepackage{color}

\newcommand{\ba}{\begin{eqnarray}}
\newcommand{\ea}{\end{eqnarray}}
\newcommand{\be}{\begin{equation}}
\newcommand{\ee}{\end{equation}}
\newcommand{\bdisplay}{\begin{displaymath}}
\newcommand{\edisplay}{\end{displaymath}}

\newcommand{\eq}[1]{Eq.\,(\ref{#1})}

\makeatletter
\def\eqnarray{\stepcounter{equation}\let\@currentlabel=\theequation
\global\@eqnswtrue
\tabskip\@centering\let\\=\@eqncr
$$\halign to \displaywidth\bgroup\hfil\global\@eqcnt\z@
  $\displaystyle\tabskip\z@{##}$&\global\@eqcnt\@ne
  \hfil$\displaystyle{{}##{}}$\hfil
  &\global\@eqcnt\tw@ $\displaystyle{##}$\hfil
  \tabskip\@centering&\llap{##}\tabskip\z@\cr}

\def\endeqnarray{\@@eqncr\egroup
      \global\advance\c@equation\m@ne$$\global\@ignoretrue}

\def\@yeqncr{\@ifnextchar [{\@xeqncr}{\@xeqncr[5pt]}}
\makeatother

\begin{document}

\title{Coulomb-nuclear interference in proton-proton scattering: Reply to a comment of V.\ Petrov }

\author{Loyal Durand}
\email{ldurandiii@comcast.net}
\altaffiliation{Mailing address: 415 Pearl Ct., Aspen, CO 81611}
\affiliation{Department of Physics, University of Wisconsin, Madison, WI 53706}
\author{Phuoc Ha}
\email{pdha@towson.edu}
\affiliation{Department of Physics, Astronomy and Geosciences, Towson University, Towson, MD 21252}

\begin{abstract}
We show that the remarks of V.\ Petrov in the preceding Comment are misdirected in the region of very small momentum transfers relevant to the study of Coulomb-nuclear interference in proton-proton scattering, and clarify the connection between our results and those of Cahn and of  Kundr\'{a}t and Lokaji\v{c}ek which he cites.
\end{abstract}


\maketitle


The strong-interaction or nuclear part of the $pp$ scattering amplitude including the effects of the Coulomb interaction and electromagnetic form-factor corrections is given in the additive eikonal model in Eq.\ (27) of \cite{DH_CoulNucl} as
\be
\label{fNc_defined}
f_{N,c}(s,q^2) =  +i\int_0^\infty db\,b\,e^{2i\delta_c(b,s)+2i\delta^{FF}_c(b,s)}\left(1-e^{2i\delta_N(b,s)}\right)J_0(qb).
\ee
Here $\delta_N=\delta_{N,R}+i\delta_{N,I}$ is the complex nuclear phase shift, $\delta_c(b,s)=\alpha\left(\log{p(W)b}+\gamma\right)$ with $\gamma$ Euler's constant is the Coulomb phase shift, and $\delta_c^{FF}(b,s)$ is the phase shift associated with the form-factor corrections.  The last is  given by a sum of hyperbolic Bessel functions with an overall factor of $\alpha$ for the standard dipole form factor $F_Q(q^2)=\mu^4/(q^2+\mu^2)^2$.

It was shown  in \cite{DH_CoulNucl}  by direct numerical calculation using the successful eikonal model in \cite{eikonal2015} that the ratio  $\lvert f_{N,c}(s,q^2)/f_N(s,q^2)\rvert$ of  the  magnitudes of the corrected amplitude to the  pure nuclear amplitude
\be
\label{f_N}
f_N(s,q^2) = i\int_0^\infty db\,b\left(1-e^{2i\delta_N(b,s)}\right)J_0(qb)
\ee
 was equal 1 to better than a part per thousand in the region $q^2\lesssim 0.15$ GeV$^2$ at 13000 GeV, and to higher $q^2$ at lower energies, covering the regions important for the determination of $\rho(s)={\rm Re} f(s,0)/{\rm Im} f(s,0)$ from Coulomb-nuclear interference effects. See Fig.\ 1 in \cite{DH_CoulNucl}. This correction is significantly smaller than the experimental uncertainties in the most accurate data available at present, and does not affect the determination of $\rho$ from 10 GeV to 13000 GeV or somewhat above. 
 
 Since our focus was on the interference effects at small $q^2$, this very small correction to the ratio of magnitudes was dropped in  \cite{DH_CoulNucl} as noted preceding Eq.\ (29), and $f_{N,c}(s,q^2)$ was taken as
\be
\label{phase_approx}
f_{N,c}(s,q^2) \approx e^{\Delta\Phi(s,q^2)}f_N(s,q^2)
\ee
in the remainder of the analysis, where $\Delta\Phi$ is the difference between the phases of $f_{N,c}$ and $f_N$,
\be
\label{phase_diff}
\Delta\Phi_N(s,q^2) = {\rm arg}f_{N,c}(s,q^2)-{\rm arg}f_N(s,q^2)={\rm arg}\left(f_{N,c}(s,q^2)/f_N(s,q^2)\right).
\ee
 The fact that the magnitude correction was dropped is the origin of the apparent discrepancy between the results in \cite{DH_CoulNucl}
and those of Cahn \cite{Cahn} and of Kundr\'{a}t and  Lokaji\v{c}ek \cite{KL-Coulomb} noted by Petrov \cite{Petrov_mag}. The correction must of course be included at large $q^2$ as is evident from the numerical results Fig.\ 1 in \cite{DH_CoulNucl}.

It is straightforward to obtain the correction to the magnitude of $f_{N,c}(s,q^2)$ relative to $f_N(s,q^2)$ at small $q^2$. We note first that the momentum-dependent part of the phase $\ln{p(W)b+\gamma}$ in $\exp{(2i\delta_c(b,s))}$ cancels in $\lvert f_{N,c}(s,q^2)\rvert^2$ . We will therefore define a modified momentum-independent phase $\delta'_c(b,s)=\alpha\log{(b/b_{\rm peak})}$, where $b_{\rm peak}$ is the value of $b$ at the peak in the impact-parameter distribution of $\lvert\left(1-e^{2i\delta_N(b,s)}\right)\rvert^2$, essentially the peak in the distribution for the imaginary part of  $f_N$. A reasonable estimate is $b_{\rm peak}\approx\sqrt{\sigma_{tot}/4\pi}$, where $\sigma_{tot}$ is the total cross section \cite{eikonal2015}. The exact result for $\lvert f_{N,c}\rvert^2$ is, of course, independent of $b_{\rm peak}$. Its use here gives a convenient way of estimating the corrections since $\delta'_c(b_{\rm peak},s)=0$  and $\delta_c^{FF}(b_{\rm peak},s)$  is also small.

With this specification, and $\delta'=\delta'_c+\delta_c^{FF}$,
\be
\label{deltaF}
\lvert f_{N,c}(s,q^2)\rvert = \lvert f_N(s,q^2)\rvert\left(1+\frac{{\rm Re}  f_N(s,q^2)\,{\rm Re} \Delta f_{N,c}(s,q^2)+{\rm Im} f_N(s,q^2)\,{\rm Im}\Delta f_{N,c}(s,q^2)}{\lvert f_N(s,q^2)\rvert^2}+\cdots\right),
\ee
where
\ba
\label{Deltaf}
\Delta f_{N,c}(s,q^2) &=&- i\int_0^\infty db\,b\left(1-e^{2i\delta'(b,s)}\right)\left(1-e^{2i\delta_N(b,s)}\right)J_0(qb) \\
\label{Deltaf2}
&\approx& -\int_0^\infty db\,b\,2\delta'(b,s)\left(1-e^{2i\delta_N(b,s)}\right)J_0(qb)+\cdots.
\ea
The expansion is justified because $2\delta'$ is small, of order $\alpha$, and the remaining factor in the integrand is compact in $b$, a major advantage of the impact-parameter description of the scattering amplitude as used in \cite{DH_CoulNucl}. This gives the corrections to leading order in $\alpha$ as
\ba
\label{Re_deltaF}
{\rm Re} \Delta f_{N,c}(s,q^2) &=& -\int_0^\infty db\,b\,2\delta'(b,s)\left(1-\cos{2\delta_{N,R}}\,e^{-2\delta_{N,I}}\right)J_0(qb), \\
\label{Im_deltaF}
{\rm Im}\Delta f_{N,c}(s,q^2) &=& \int_0^\infty db\,b\,2\delta'(b,s)\sin{2\delta_{N,R}}\,e^{-2\delta_{N,I}}J_0(qb).
\ea

The factors in the integrals for ${\rm Re}\Delta f_{N,c}$ and ${\rm Im}\Delta f_{N,c}$ other than  $2\delta'(b,s)$ are just the integrands for ${\rm Im} f_N$ and ${\rm Re} f_N$, respectively. The integrands for the latter quantities are fairly sharply peaked near $b=b_{\rm peak}$ for ${\rm Im} f_N$ at $q^2=0$, and slightly beyond for ${\rm Re} f_N$. If the phase shifts were constant over that region, they would factor out of the integrals and we would have ${\rm Re}\Delta f_{N,c}  \approx -2\delta'{\rm Im}f_N $ and ${\rm Im}\Delta f_{N,c}  \approx 2\delta'{\rm Re}f_N $ and the correction to $\lvert f_N\rvert$ would vanish. Because of the variation in $\delta'(b,s)$ over the peak regions and the small difference in the locations of the peaks in the integrands, the cancellation is only approximate. Given the structure of the correction, it is nevertheless clear that the correction is at most of order $\alpha\rho(s)\sim 10^{-3}$; it is smaller in the exact calculations. This estimate extends over the region in $q^2$ in which the variation of the Bessel function $J_0(qb)=1-\frac{1}{4}(qb)^2+\cdots$ can be ignored, $q^2\lesssim 4/b_{\rm peak}^2$, {\em e.g.\ }$q^2\lesssim 0.12$ GeV$^2$ at 13000 GeV. This is in agreement with the exact results in Fig.\ 1 of \cite{DH_CoulNucl}, where the magnitude of the correction begins to increase significantly beyond that point.

The expression in \eq{Deltaf} is the starting point for the treatment of the Coulomb effects by Cahn (\cite{Cahn}, Eq.\ (15)), with the form-factor effects included later in Eq.\ (28). This treatment was later sharpened by Kundr\'{a}t and  Lokaji\v{c}ek \cite{KL-Coulomb}. Cahn uses the Fourier convolution theorem to write the integral as the convolution of the Fourier transforms of the two factors multiplying the Bessel function in \eq{Deltaf}. To obtain convergence of the transform of the factor $(1-e^{2i\delta'})$, he first replaces the Coulomb phase shift by that for a massive photon, $\alpha/q^2\rightarrow \alpha/(q^2+\lambda^2)$, and then rearranges the terms in the full expression into a form in which he can take the limit $\lambda\rightarrow 0$, with the final result given in his Eq.\ (30). This gives
\be
\label{Cahn_form}
\Delta f_{N,c}(s,q^2) = -\frac{i}{\pi}\int d^2k\frac{2\alpha}{({\bf k}-{\bf q})^2}F_Q^2(({\bf k}-{\bf q})^2)\left[f_N(s,k^2)-f_N(s,q^2)\right]
\ee
with our notation and normalization.

In this approach, the expression for the fractional change in $\lvert f_{N,c}\rvert^2$ is given by 
\be
\label{delta_fmag}
\delta\lvert f_{N,c}\rvert = \frac{1}{\pi}\frac{1}{\lvert f_N(s,q^2)\rvert^2}\int d^2k\frac{2\alpha}{({\bf k}-{\bf q})^2}F_Q^2(({\bf k}-{\bf q})^2)\left[{\rm Re}f_N(s,q^2){\rm Im}f_N(s,k^2)-{\rm Im}f_N(s,q^2){\rm Re}f_N(s,k^2)\right].
\ee
This expression has the same structure as \eq{Deltaf}, is nonsingular, of order $\alpha\rho(s)$. It would only vanish identically for the ratio ${\rm Re}f_N(s,k^2)/{\rm Im}f_N(s,k^2)$ constant and equal to ${\rm Re}f_N(s,q^2)/{\rm Im}f_N(s,q^2)$. This would require a constant nuclear phase as observed by Petrov \cite{Petrov_mag}. While correct, this remark is not relevant to the treatment of Coulomb-nuclear interference in \cite{DH_CoulNucl}, where the correction was only omitted in a region in which it was shown to be negligibly small, and not elsewhere.

In the eikonal model in \cite{eikonal2015} and other models which respect the constraints on the phase of $f_N$ imposed by unitarity and analyticity  \cite{Blockreview}, ${\rm Re}f_N$ actually decreases  at small $k^2$ (or $q^2$) more rapidly than ${\rm Im}f_N$  as $k^2$ ($q^2$) increases,  so the cancellation in \eq{delta_fmag} is not complete. The result is consistent with that obtained above, where the shift of the impact-parameter distribution for ${\rm Re} f_N$ toward larger $b$ than that for ${\rm Im}f_N$ leads to the more rapid decrease of  ${\rm Re} f_N$ through the earlier onset of the effects of the Bessel function in \eq{f_N}. 

The results in Eqs.\ (\ref{deltaF}) and (\ref{delta_fmag}) are equivalent, and the corrections to $\lvert f_{N,c}\rvert $ are very small, {\em e.g.}, about $6\times 10^{-4}$ at $q^2=0$ for either from 100 to 13000 GeV, and of similar size throughout the region $q^2\lesssim 4/b_{\rm peak}^2$ in \cite{DH_CoulNucl}. We conclude that there is no problem in the small-$q^2$ region in which the results on Coulomb-nuclear interference in \cite{DH_CoulNucl} were derived with the magnitude correction neglected, contrary to Petrov's comments \cite{Petrov_mag}. The phase of the nuclear amplitude is not constant as shown in the exact calculations. The ratio of magnitudes $\lvert f_{N,c}(s,q^2)/f_N(s,q^2)\rvert$ is 1 to high accuracy and the correction can be negected  for $q^2\lesssim 4/b_{\rm peak}^2$ where the Coulomb-nuclear interference effects are significant.  This changes at larger $q^2$ as shown in \cite{DH_CoulNucl} as the diffractive structure of the nuclear amplitude becomes evident.

\begin{acknowledgments}
We would like to thank Dr.\ V.\ Petrov for raising the issues treated here, and for lively correspondence about them.
L.D.  would  like to thank the Aspen Center for Physics for its hospitality and for its partial support of this work under NSF Grant No. 1066293.  P.H.\ would like to thank Towson University Fisher College of Science and Mathematics for support.
\end{acknowledgments}

\bibliography{CoulombMathbib}

\begin{thebibliography}{6}%
\makeatletter
\providecommand \@ifxundefined [1]{%
 \@ifx{#1\undefined}
}%
\providecommand \@ifnum [1]{%
 \ifnum #1\expandafter \@firstoftwo
 \else \expandafter \@secondoftwo
 \fi
}%
\providecommand \@ifx [1]{%
 \ifx #1\expandafter \@firstoftwo
 \else \expandafter \@secondoftwo
 \fi
}%
\providecommand \natexlab [1]{#1}%
\providecommand \enquote  [1]{``#1''}%
\providecommand \bibnamefont  [1]{#1}%
\providecommand \bibfnamefont [1]{#1}%
\providecommand \citenamefont [1]{#1}%
\providecommand \href@noop [0]{\@secondoftwo}%
\providecommand \href [0]{\begingroup \@sanitize@url \@href}%
\providecommand \@href[1]{\@@startlink{#1}\@@href}%
\providecommand \@@href[1]{\endgroup#1\@@endlink}%
\providecommand \@sanitize@url [0]{\catcode `\\12\catcode `\$12\catcode
  `\&12\catcode `\#12\catcode `\^12\catcode `\_12\catcode `\%12\relax}%
\providecommand \@@startlink[1]{}%
\providecommand \@@endlink[0]{}%
\providecommand \url  [0]{\begingroup\@sanitize@url \@url }%
\providecommand \@url [1]{\endgroup\@href {#1}{\urlprefix }}%
\providecommand \urlprefix  [0]{URL }%
\providecommand \Eprint [0]{\href }%
\providecommand \doibase [0]{http://dx.doi.org/}%
\providecommand \selectlanguage [0]{\@gobble}%
\providecommand \bibinfo  [0]{\@secondoftwo}%
\providecommand \bibfield  [0]{\@secondoftwo}%
\providecommand \translation [1]{[#1]}%
\providecommand \BibitemOpen [0]{}%
\providecommand \bibitemStop [0]{}%
\providecommand \bibitemNoStop [0]{.\EOS\space}%
\providecommand \EOS [0]{\spacefactor3000\relax}%
\providecommand \BibitemShut  [1]{\csname bibitem#1\endcsname}%
\let\auto@bib@innerbib\@empty
\bibitem [{\citenamefont {Durand}\ and\ \citenamefont
  {Ha}(2020)}]{DH_CoulNucl}%
  \BibitemOpen
  \bibfield  {author} {\bibinfo {author} {\bibfnamefont {L.}~\bibnamefont
  {Durand}}\ and\ \bibinfo {author} {\bibfnamefont {P.}~\bibnamefont {Ha}},\
  }\href@noop {} {\bibfield  {journal} {\bibinfo  {journal} {Phys. Rev. D}\
  }\textbf {\bibinfo {volume} {102}},\ \bibinfo {pages} {036025} (\bibinfo
  {year} {2020})},\ \Eprint {http://arxiv.org/abs/arXiv:2007.07827v3 [hep-ph]}
  {arXiv:2007.07827v3 [hep-ph]} \BibitemShut {NoStop}%
\bibitem [{\citenamefont {Block}\ \emph {et~al.}(2015)\citenamefont {Block},
  \citenamefont {Durand}, \citenamefont {Ha},\ and\ \citenamefont
  {Halzen}}]{eikonal2015}%
  \BibitemOpen
  \bibfield  {author} {\bibinfo {author} {\bibfnamefont {M.~M.}\ \bibnamefont
  {Block}}, \bibinfo {author} {\bibfnamefont {L.}~\bibnamefont {Durand}},
  \bibinfo {author} {\bibfnamefont {P.}~\bibnamefont {Ha}}, \ and\ \bibinfo
  {author} {\bibfnamefont {F.}~\bibnamefont {Halzen}},\ }\href@noop {}
  {\bibfield  {journal} {\bibinfo  {journal} {Phys.\ Rev.\ D}\ }\textbf
  {\bibinfo {volume} {92}},\ \bibinfo {pages} {014030} (\bibinfo {year}
  {2015})},\ \Eprint {http://arxiv.org/abs/arXiv:1505.04842v1 [hep-ph]}
  {arXiv:1505.04842v1 [hep-ph]} \BibitemShut {NoStop}%
\bibitem [{\citenamefont {Cahn}(1982)}]{Cahn}%
  \BibitemOpen
  \bibfield  {author} {\bibinfo {author} {\bibfnamefont {R.}~\bibnamefont
  {Cahn}},\ }\href@noop {} {\bibfield  {journal} {\bibinfo  {journal} {Z. Phys.
  C}\ }\textbf {\bibinfo {volume} {15}},\ \bibinfo {pages} {253} (\bibinfo
  {year} {1982})}\BibitemShut {NoStop}%
\bibitem [{\citenamefont {Kundr\'{a}t}\ and\ \citenamefont
  {Lokaji\v{c}ek}(1994)}]{KL-Coulomb}%
  \BibitemOpen
  \bibfield  {author} {\bibinfo {author} {\bibfnamefont {V.}~\bibnamefont
  {Kundr\'{a}t}}\ and\ \bibinfo {author} {\bibfnamefont {M.}~\bibnamefont
  {Lokaji\v{c}ek}},\ }\href@noop {} {\bibfield  {journal} {\bibinfo  {journal}
  {Z. Phys. C}\ }\textbf {\bibinfo {volume} {63}},\ \bibinfo {pages} {619}
  (\bibinfo {year} {1994})}\BibitemShut {NoStop}%
\bibitem [{\citenamefont {Petrov}(2020)}]{Petrov_mag}%
  \BibitemOpen
  \bibfield  {author} {\bibinfo {author} {\bibfnamefont {V.~A.}\ \bibnamefont
  {Petrov}},\ }\href@noop {} {\  (\bibinfo {year} {2020})},\ \Eprint
  {http://arxiv.org/abs/arXiv:2008.00287 [hep-ph]} {arXiv:2008.00287 [hep-ph]}
  \BibitemShut {NoStop}%
\bibitem [{\citenamefont {Block}(2006)}]{Blockreview}%
  \BibitemOpen
  \bibfield  {author} {\bibinfo {author} {\bibfnamefont {M.~M.}\ \bibnamefont
  {Block}},\ }\href@noop {} {\bibfield  {journal} {\bibinfo  {journal} {Physics
  Reports}\ }\textbf {\bibinfo {volume} {436}},\ \bibinfo {pages} {71}
  (\bibinfo {year} {2006})}\BibitemShut {NoStop}%
\end{thebibliography}%

\end{document}